\documentclass[twocolumn]{aastex61}
\shorttitle{Numerical Simulations of SNR Evolution in a Cloudy ISM}
\shortauthors{Slavin et al.}
\newcommand{\cc}{cm$^{-3}$}
\begin{document}

\title{Numerical Simulations of Supernova Remnant Evolution in a Cloudy
Interstellar Medium}

\correspondingauthor{Jonathan D. Slavin}
\email{jslavin@cfa.harvard.edu}

\author[0000-0002-7597-6935]{Jonathan D. Slavin}
\affiliation{Harvard-Smithsonian Center for Astrophysics, 60 Garden Street,
Cambridge, MA 02138, USA}

\author[0000-0003-4284-4167]{Randall K. Smith}
\affiliation{Harvard-Smithsonian Center for Astrophysics, 60 Garden Street,
Cambridge, MA 02138, USA}

\author[0000-0003-3462-8886]{Adam Foster}
\affil{Harvard-Smithsonian Center for Astrophysics, 60 Garden Street,
Cambridge, MA 02138, USA}

\author[0000-0002-6678-590X]{Henry D. Winter}
\affil{Harvard-Smithsonian Center for Astrophysics, 60 Garden Street,
Cambridge, MA 02138, USA}

\author[0000-0002-7868-1622]{John C. Raymond} 
\affil{Harvard-Smithsonian Center for Astrophysics, 60 Garden Street,
Cambridge, MA 02138, USA}

\author[0000-0002-6986-6756]{Patrick O. Slane}
\affil{Harvard-Smithsonian Center for Astrophysics, 60 Garden Street,
Cambridge, MA 02138, USA}

\author[0000-0002-5092-6085]{Hiroya Yamaguchi}
\affiliation{NASA Goddard Space Flight Center, Code 662, Greenbelt, MD 20771,
    USA}
\altaffiliation{Department of Astronomy, University of Maryland, College
    Park, MD 20742, USA}


\begin{abstract}
    The mixed morphology class of supernova remnants has centrally peaked
    X-ray emission along with a shell-like morphology in radio emission.
    White \& Long proposed that these remnants are evolving in a cloudy medium
    wherein the clouds are evaporated via thermal conduction once being
    overrun by the expanding shock.  Their analytical model made detailed
    predictions regarding temperature, density and emission profiles as well
    as shock evolution.  We present numerical hydrodynamical models in 2D and
    3D including thermal conduction, testing the White \& Long model and
    presenting results for the evolution and emission from remnants evolving
    in a cloudy medium.  We find that, while certain general results of the
    White \& Long model hold, such as the way the remnants expand and the
    flattening of the X-ray surface brightness distribution, in detail there
    are substantial differences.  In particular we find that the X-ray
    luminosity is dominated by emission from shocked cloud gas early on,
    leading to a bright peak which then declines and flattens as evaporation
    becomes more important.  In addition, the effects of thermal conduction on
    the intercloud gas, which is not included in the White \& Long model, are
    important and lead to further flattening of the X-ray brightness profile
    as well as lower X-ray emission temperatures.
\end{abstract}
\keywords{hydrodynamics, ISM: clouds, ISM: supernova remnants, X-rays: ISM}

\section{Introduction}
Supernovae are the primary source of hot gas in galaxies.  The shocks created
in a supernova explosion can expand to distances of more than 100 pc and
create large, hot, low density volumes of gas that persist for millions of
years.  Supernova remnant (SNR) evolution has been explored theoretically for
several decades with increasing levels of complexity.
However, the appearance of SNRs and the way the shocks couple with the
interstellar medium (ISM) depend crucially on the nature of the medium into
which the remnant expands, in particular its density and inhomogeneity.

For many years SNRs were divided into two categories, shell-like and
plerionic.  The first category has emission that peaks at the edge of the
remnant while plerions are centrally brightened.  It was believed that in the
latter case a pulsar wind powers the emission, while for shell-like remnants,
emission from the outer blastwave shock dominates. However, cases that did not
fit either morphology were found and referred to as either thermal composite,
because of their centrally concentrated and yet thermal X-ray emission along
with edge brightened radio emission, or mixed morphology supernova remnants
(MMSNR).

\citet{Rho+Petre_1998} first argued that the mixed morphology remnants formed
a truly distinct class of SNRs.  In more recent years more remnants in this
class have been found and evidence has been presented that these remnants are,
as a rule, undergoing interactions with dense clouds.  In particular
\citet{Yusef-Zadeh_etal_2003} showed a strong correlation of OH maser emission
with mixed morphology, indicating that shocks in molecular clouds are present
around MMSNRs.  In addition, shock-cloud interaction indicators
such as broad velocity wings \citep{Frail+Mitchell_1998} or an enhanced
$^{12}\mathrm{CO}(J=2-1)/^{12}\mathrm{CO}(J=1-0)$ ratio \citep{Seta_etal_1998}
are observed in many MMSNRs \citep[see review by][]{Slane_etal_2015}.

At present, there are $\sim 40$ known MMSNRs, but the nature of the central
X-ray emission is poorly understood. X-ray spectra show evidence for the
presence of ejecta in some such MMSNRs, but the inferred mass of the
X-ray emitting material appears to be much larger than any reasonable ejecta
contributions. The observed temperature profiles show little variation with
radius, in contrast to the steep profile associated with the Sedov phase.
Possibly related is the observation that the plasma in many MMSNRs is observed
to be overionized \citep[e.g.,][]{Kawasaki_etal_2002, Yamaguchi_etal_2009},
indicating a rapid cooling phase \citep{Uchida_etal_2012, Moriya_2012}.

Nearly half of the known MMSNRs are observed to produce gamma-ray emission,
presumably indicating the interaction between protons accelerated by the SNR
(or re-accelerated cosmic rays) and molecular cloud material or dense
post-shock regions of radiative shocks \citep{Uchiyama_etal_2010,
Lee_etal_2015}.

Even well before many of these observations that characterized MMSNRs in
detail, \citet[hereafter WL]{White+Long_1991} put forward an analytical model
that attempted to explain remnants with centrally peaked X-ray emission.  WL
included terms in the fluid equations describing evolution in spherical
symmetry of a SNR that allowed for a continual injection of mass and momentum
as the result of cloud evaporation.  In this way they aimed to describe the
evolution and emission distribution generated when a SN explodes in a cloudy
medium and where thermal conduction evaporates the clouds that have been
enveloped by the blastwave.  We note that there was no term in the equations
accounting for the transfer of energy within the hot gas by thermal
conduction.  We will discuss the ramifications of this further below.

\section{Methods} \label{sec:meth}
WL created their models for SNR evolution in a cloudy medium starting from the
usual set of equations for hydrodynamics under the assumption of spherical
symmetry but with added terms to account for the effects of cloud evaporation
on the mass, momentum and energy in the remnant. Though they included a term
for mass injection into the intercloud medium as a result of thermal
conduction, as mentioned above, they did not explicitly include thermal
conduction. For our numerical hydrodynamical calculations presented here we
employ the code FLASH
\citep[\url{http://flash.uchicago.edu/site/flashcode/}]{Fryxell_etal_2000,
Fryxell_etal_2010} without any extra terms added to account for mass loss from
clouds since they are unnecessary in multiple dimensions as long as the
physics is correctly modeled.  An essential aspect of the modeling is the
inclusion of thermal conduction, which is accurately and efficiently included
via the \texttt{Diffuse} module included in version 4.3 of FLASH.  We discuss
the details of our use of this module in the appendix.  The initial conditions
for our cloudy medium runs include an initial distribution of clouds in space
with varying sizes. The cloud size distribution and density were not important
for WL as long as the fundamental assumptions of their calculations were not
violated, namely that the clouds are much denser than the intercloud medium,
the clouds are uniformly distributed in space and the filling factor of the
clouds is small.  We fulfill these criteria by randomly placing clouds that
are a factor of 100 times denser than the intercloud medium with an overall
small volume filling factor (dictated by the $C$ parameter, see below).  We
set off the SN explosion in this medium by putting the appropriate amount of
thermal energy in a small number of parcels at the center of the remnant. We
do not include any ejecta mass, which would be important at early times
during the ejecta dominated phase. The characteristic time for
the transition from the ejecta dominated to Sedov phase is
\citep{Truelove+McKee_1999} $t_{ch} = E^{-1/2} M_\mathrm{ej} \rho^{-1/3}$,
which for our values of explosion energy and density, and assuming 5 $M_\odot$
of ejecta, is about 2700 yr. It will take several times this value for the
evolution to truly settle into a Sedov solution evolution in the remnant as a
whole, though the approach will be faster for the region behind the forward
shock. For this reason we do not expect the early time evolution (e.g. X-ray
luminosity) to be very realistic (though of course this also applies to
the WL model). In our runs the thermal energy leads to rapid
expansion and partial conversion of the thermal energy to kinetic energy as
expected for Sedov-Taylor expansion \citep[e.g.,][]{Chevalier_1974}.  Most of
the results we discuss in this paper were carried out in 2D cylindrical
symmetry, though we have done runs in 3D as well as a check on the effects of
the 2D assumption.

For all of our runs, we use an explosion energy of $10^{51}$ ergs.  As with
the Sedov-Taylor solution, we expect the explosion energy to set the overall
scale of the remnant but not to affect such things as the division of energy
between thermal and kinetic forms.  We use an ambient intercloud density of
0.25 cm$^{-3}$ and a temperature of $10^4$ K.  These values are typical for
the warm interstellar medium, though the temperature is perhaps slightly high
for realistic heating and cooling balance.  The exact value for the
temperature should not have a significant affect on our results.  The clouds
are assumed to have a density of 25 cm$^{-3}$, which then leads to a
temperature of 100 K to maintain pressure balance in the ambient medium.
(Note that we are not tracking the ionization which would change these values
if the ionization level were lower in the clouds than in the intercloud
medium.)  These values are consistent with typical cold neutral (\ion{H}{1})
cloud values.  We do not expect the cloud temperature to be important to the
remnant evolution.  The cloud density could have some effect, since a larger
density, for the same ratio of cloud mass to intercloud mass, would imply a
lower cloud filling factor, though we think that our assumed density is
realistic. For the cloud sizes we use a power law in radius.  We took the
distribution used in \citet{McKee+Ostriker_1977}, which was partly based on
observations by \citet{Hobbs_1974} and has lower and upper radius cutoffs of
0.48 and 10 pc and a power law exponent of -3 in radius.  We create
realizations of the cloud distribution by drawing randomly from this power law
distribution and also choosing cloud location randomly within the simulation
volume. We discard clouds that overlap with previously generated clouds such
that each cloud is distinct from all the others.  We also discard clouds that
extend outside of the simulation volume.  For our 2D cylindrically symmetric
runs the clouds are, in effect, tori with circular cross sections.  In our 3D
runs, which are done in cartesian coordinates, the clouds are true spheres.

The actual cloud size distribution in the Galaxy is poorly constrained and so
our adopted parameters of the cloud size distribution may not accurately
reflect the true cloud size distribution, \citep[though others have found
similar results, e.g.][]{Gosachinskij+Morozova_1999,Chieze+Lazareff_1980}. 
There is evidence that interstellar clouds have a morphology that is closer to
filamentary or ``clumpy sheets'' \citep{Heiles+Troland_2003}, though more
regular clouds are observed as well.  We do not expect the cloud geometry to
have a very substantial effect on SNR evolution or observables since, once
shocked, the complex flows inside the remnant create large distortions of the
clouds in any case. However, if a remnant encounters a cloud that is
comparable in size to the remnant, both the appearance and evolution of the
remnant would be strongly affected. In such cases the remnant would have a
strongly asymmetric appearance. Exploration of such cases is beyond the scope
of this paper.

WL found that their solutions could be parametrized by just two parameters:
$C$, the ratio of the mass in clouds to that in the intercloud medium, and
$\tau$ the ratio of the cloud evaporation timescale to the remnant age.  While
$C$ is clearly a fixed value that characterizes the conditions in the ambient
medium, WL concede that $\tau$ may not be. Since the temperature of the
interior of a remnant evolves with time and the evaporation timescale would
generally be expected to depend on the temperature and/or the pressure.  In
fact it can be shown, using the results of \citet{Dalton+Balbus_1993}, that in
the limit of highly saturated evaporation, the evaporation timescale is
$\tau_\mathrm{evap} \propto R_\mathrm{cl}^{7/6} P^{-5/6}$, with no explicit
dependence on temperature. Since the pressure in the Sedov-Taylor solution
drops as $t^{-6/5}$, if the medium surrounding the enveloped clouds is close
to that for Sedov-Taylor expansion in a uniform medium, we expect that the
evaporation timescale should be proportional to the remnant age and so $\tau$
should be constant.  Deviations from these assumptions, for example because
the evaporation is not highly saturated or because the cloud-cloud or
cloud-shock interactions cause substantial deviations from a Sedov-Taylor
evolution that lead to different evolution of the evaporation rate, would then
be expected to cause variation of $\tau$ with remnant age. We discuss this
further below.

\begin{figure*}[ht!]
    \plotone{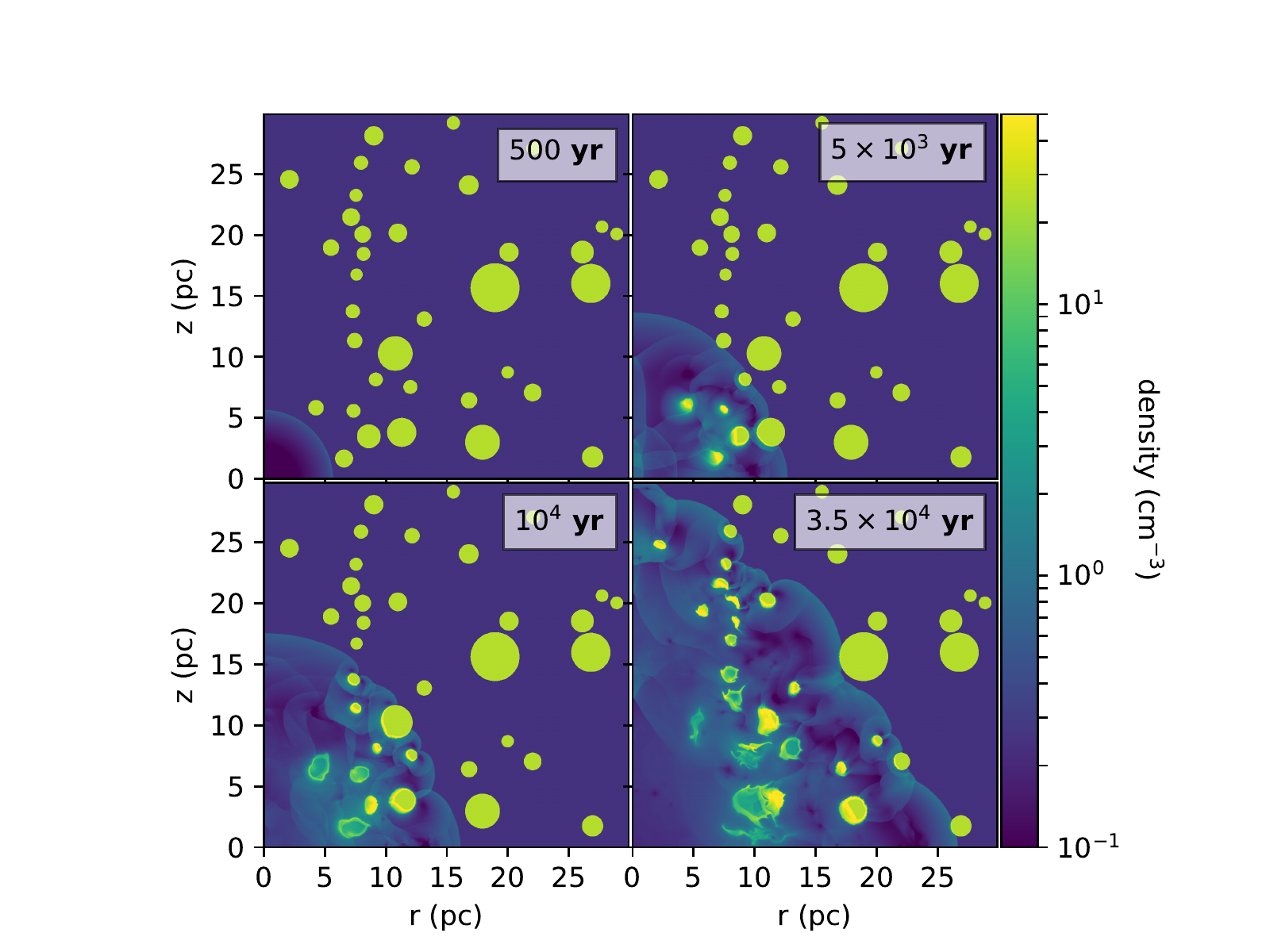}
    \caption{Snapshots of the density for a 2D cylindrically symmetric run of
        SNR evolution in a cloudy medium.  The cloud filling factor is set by
        the WL parameter $C = 10$.\label{fig:snrevol}}
\end{figure*}

\begin{figure*}[ht!]
    \plotone{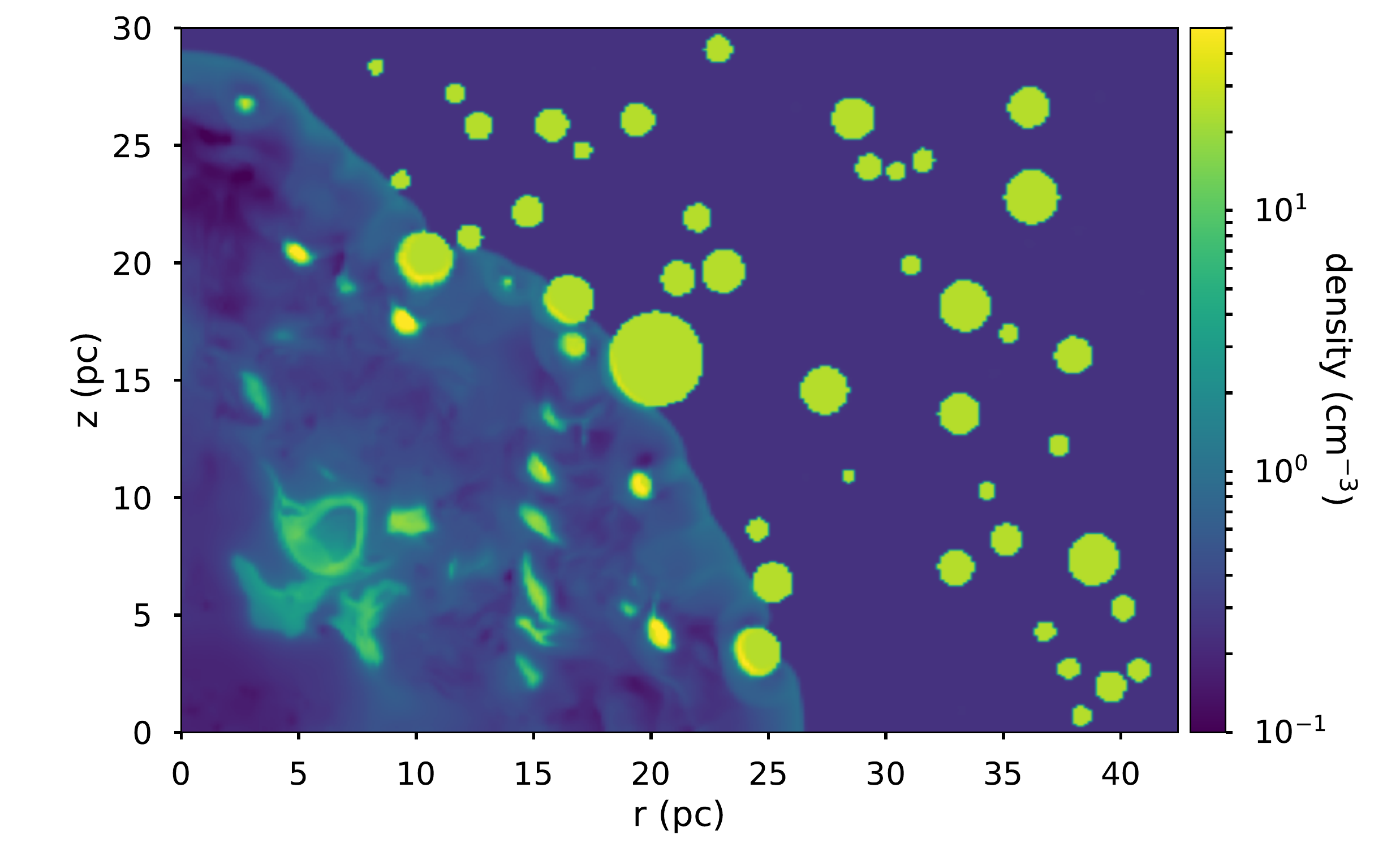}
    \caption{Slice of the density for a 3D run of SNR evolution in a cloudy
        medium, again with WL parameter $C = 10$. The slice is taken
        parallel to the $z$ axis and at a 45$\degr$ angle to the $x$ and $y$
        axes. The age of the remnant here is $3.5\times 10^4$ yr.
        \label{fig:slice3D}}
\end{figure*}

\section{Results}
In Figure \ref{fig:snrevol} we show the density for four snapshots of the
evolution for one simulation.  For this case we use a value for the WL
parameter $C = 10$, ratio of mass in clouds to that in the intercloud medium.
Given the definition of $C$, the volume filling factor of clouds is
\begin{equation}
    f = \frac{C}{\chi + C}
\end{equation}
where $\chi$ is the ratio of the cloud density to the intercloud density.
With our assumption that $\chi = 100$, for the $C = 10$ case, $f = 0.091$.
For our 2D cylindrically symmetric runs, the clouds are really toroids.
However, for these cases the area filling factor, that is the fraction of the
$r-z$ plane filled with clouds, is quite close to the volume filling factor of
the toroidal clouds. If the clouds are evenly distributed in cylindrical
radial distance this will always be the case.  We have explored a range of $C$
values ranging from 3 to 30.  We take 10 as our standard case for much of the
discussion in this paper. Figure \ref{fig:slice3D} shows the density for a
slice through one of our 3D runs.  That run also had $C = 10$.  Note the
similar area filling factor for the clouds.  The fact that it is a slice
through a 3D volume results in the cloud radii appearing systematically
smaller than the true cloud radii.

\begin{figure*}[ht!]
    \includegraphics[width=0.49\textwidth]{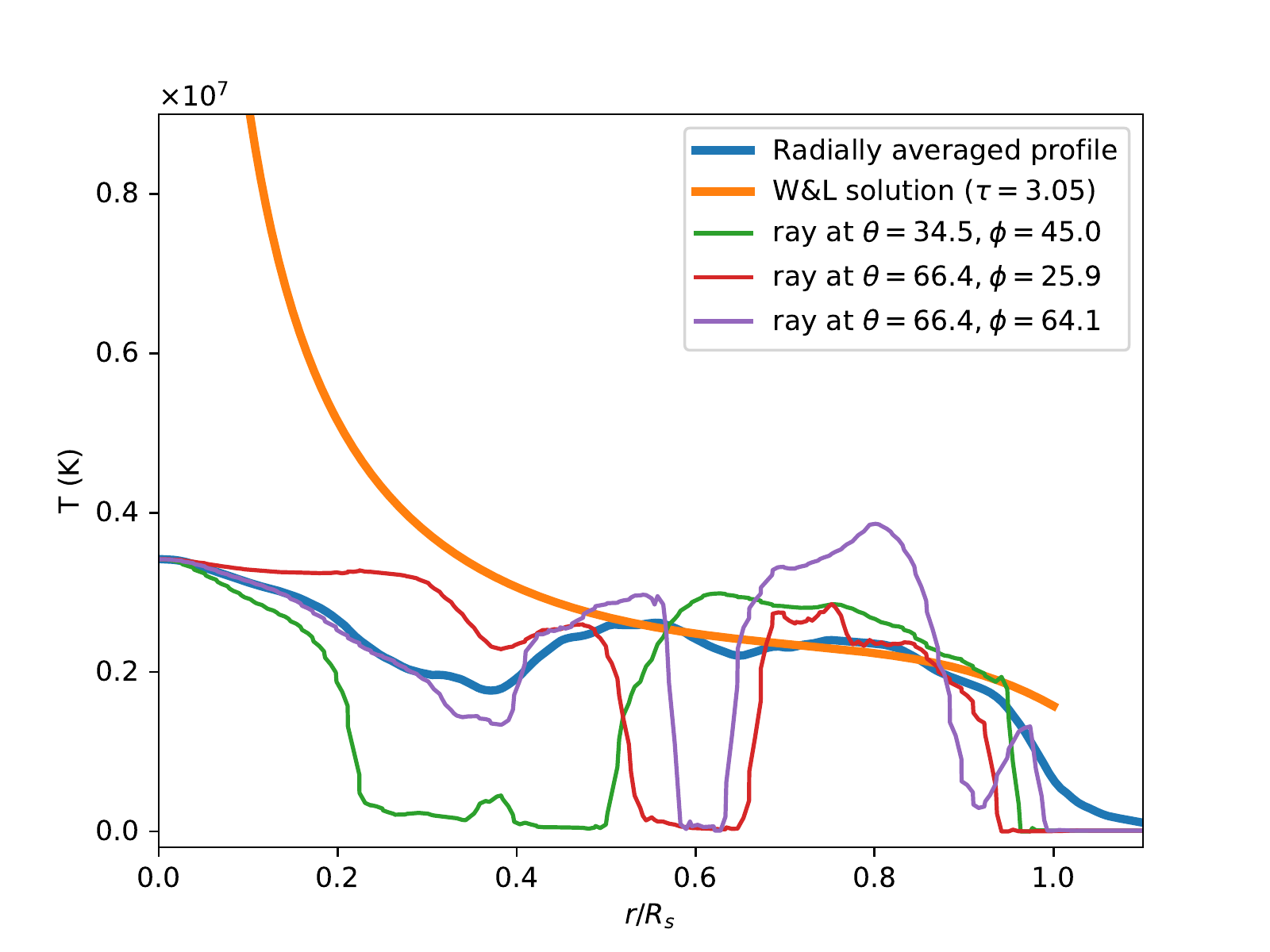}
    \includegraphics[width=0.49\textwidth]{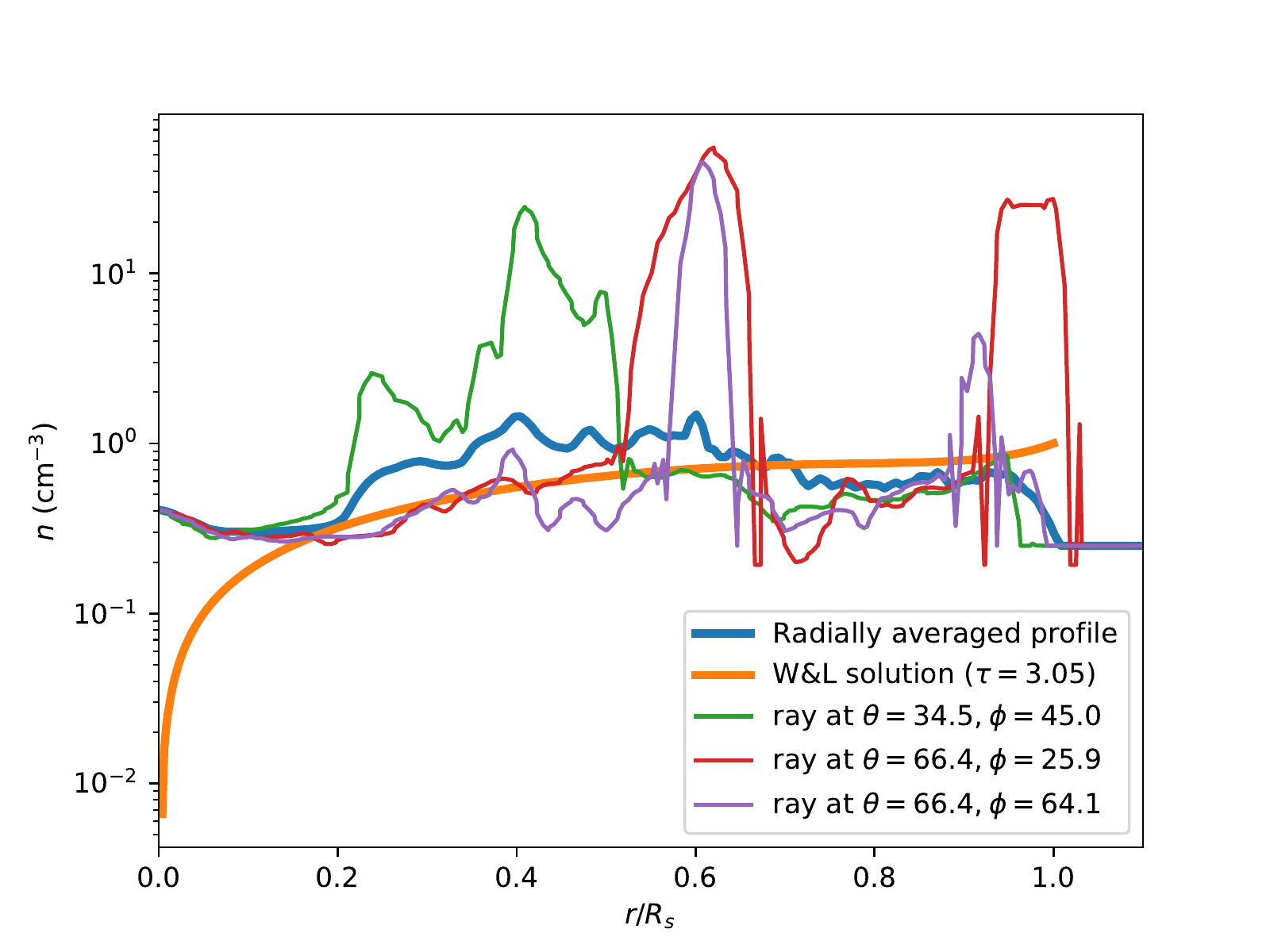}
    \caption{Temperature (\textit{left}) and density (\textit{right}) radial
        profiles for a 3D run.  The rays show all the variations along the
        particular sight line as indicated in the legend (using spherical
        coordinate) while the ``radially averaged'' curves only include the
        parcels that are hot ($T > 10^5$ K).  The ``W\&L solution'' curves
        show the WL type profiles for the actual values of $C$ in the
        simulation and the value for $\tau$ that results in the same mass in
        hot gas as in the simulation. The SNR age for this run is $2.5\times
        10^4$ yr.\label{fig:prof}}
\end{figure*}

In Figure \ref{fig:prof} we show temperature and density profiles for the same
3D run as in Figure \ref{fig:slice3D}.  The rays shown illustrate the
variations along a line of sight from the origin outward along the direction
given by the angles listed in the legend.  We also show the radially averaged
profiles for temperature and density, but for those we exclude cloud material,
here defined as parcels with temperature below $10^5$ K.  The averaging is
done by calculating a volume weighted sum of all parcels within each radial
bin that have hot gas in them and dividing by the volume of parcels (whether
or not they have hot gas) in the same radial bin.  The WL profile uses the
actual value of $C$ for the simulation at the given time ($2.5\times10^4$ yr)
and given the current shape of the shock front ($C = 9.0$), which is
calculated as the mass in clouds within the shock front divided by the mass in
the intercloud medium inside the same volume, both for the initial,
undisturbed medium.  The value of $\tau$ used is that which leads to the same
mass of hot gas as for the simulation.  We see that, while for much of the
outer parts of the remnant, the radially averaged profile is similar to the WL
profile, in the inner region the density and temperature stay flat for the
simulation while the WL profile has a steeply rising temperature and falling
density similar to a Sedov-Taylor type profile.  In addition, the density
variations, as well as averaged values, are important for the X-ray emission,
as discussed below, since emissivity goes as density squared.

\begin{figure}[ht!]
    \includegraphics[width=0.49\textwidth]{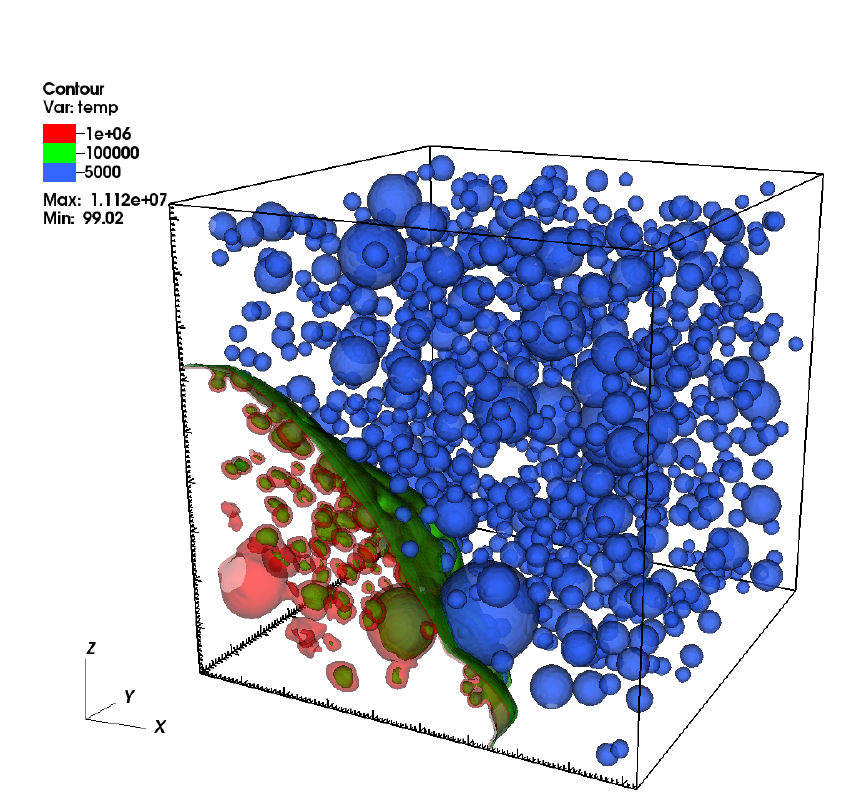}
    \caption{3D rendering of the shock and clouds for one of our 3D runs.  The
        cloud and shock surfaces are shown as temperature contours. The age of
        the remnant is $1.5\times 10^4$ yr.  The box is 30 pc on a side.  Note
        the strong distortions of the shock front.
    \label{fig:3Dshock}}
\end{figure}

\subsection{Shock Evolution}
The similarity solution of WL results in expansion of the shock front with the
same power law in time as for the Sedov-Taylor solution, i.e.\ $R_s \propto
t^{2/5}$.  For our calculations, evaluating the shock radius is not entirely
straightforward since in regions where a cloud is being encountered the shock
is slowed and the front in general is complex. In Figure \ref{fig:3Dshock} we
show 3D contours of the shock and clouds which illustrate the complexity of
the shock front.  We found that the simplest approach and the one that is
closest to what one would find from observations is to find the outermost
pressure contour that is substantially above the ambient pressure, $P/k_B =
10^4$ cm$^{-3}$ K for our case with an ambient pressure of 5200 cm$^{-3}$ K,
and use that to calculate the volume inside the remnant.  The shock radius is
then just $R_{s,\mathrm{eff}} = (3 V_s/4\pi)^{1/3}$, where $V_s$ is the volume
enclosed by the shock.  (Note: the \texttt{find\_contours} and
\texttt{grid\_points\_in\_poly} methods in the scikit-image measure python
module provide effective methods for finding parcels inside the shock.) This
leads to a relatively smooth shock expansion except for some cases in which
the shock wraps around a large cloud leading to a jump in the shock radius.
An alternative and easier approach is to use the volume of hot gas to define
the shock radius, though in that case the volume in the clouds contained in
the SNR is not included.  In practice for our calculations that volume is
small compared with the hot gas volume and the derived shock radius is very
similar.

\begin{figure}[ht!]
    \includegraphics[width=0.49\textwidth]{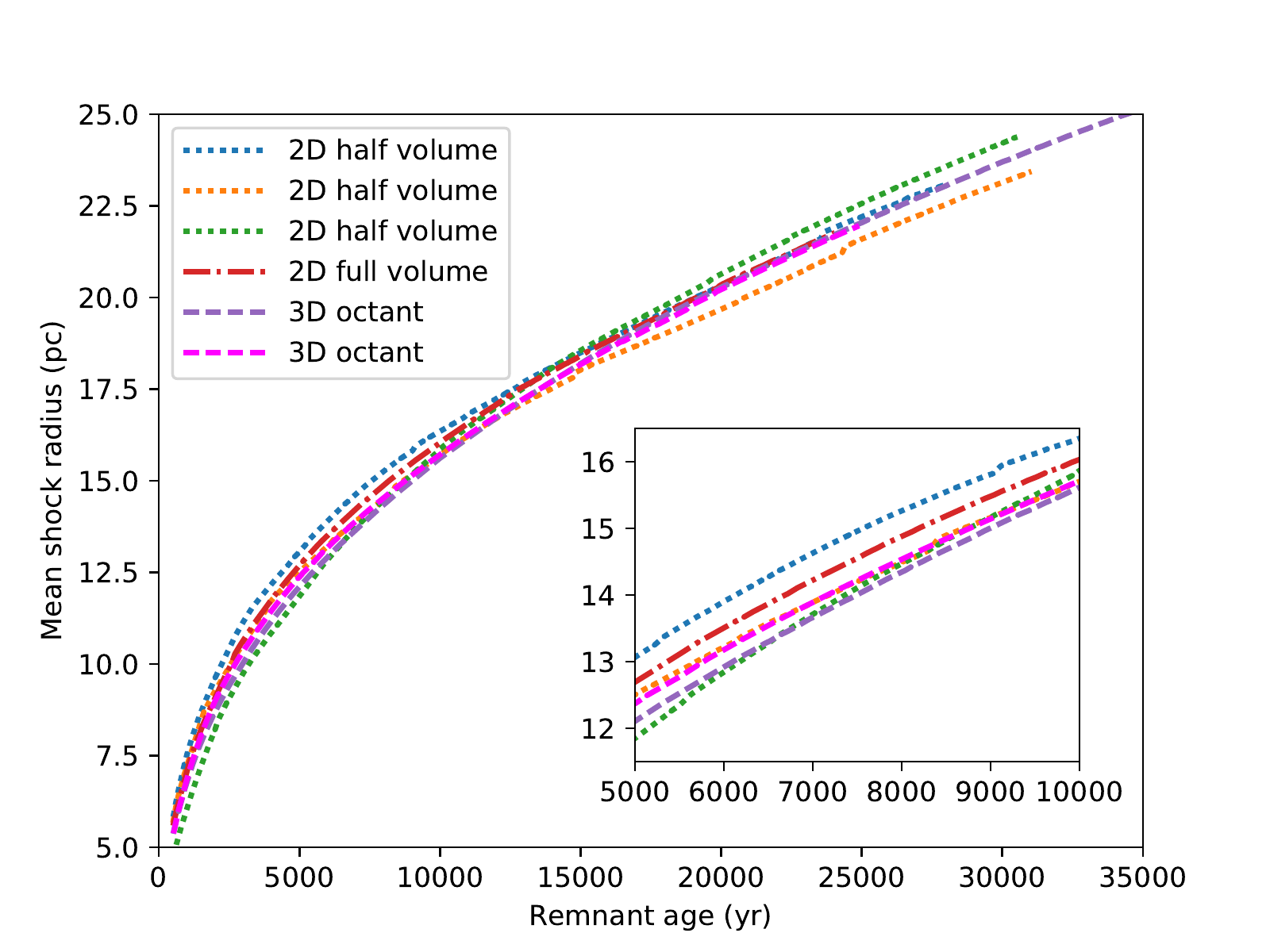}
    \caption{Evolution of the mean shock radius for different $C = 10$ runs.
        This illustrates the variations that are possible with different cloud
        realizations. We note that the shock evolutions for the 3D cases are
        consistent with the 2D cylindrically symmetric cases as is the 2D case
        using the full volume (positive and negative $z$). The inset shows a
        detail to illustrate more clearly the differences during a medium age
        period. \label{fig:rs_evol}}
\end{figure}

\begin{figure}[ht!]
    \includegraphics[width=0.49\textwidth]{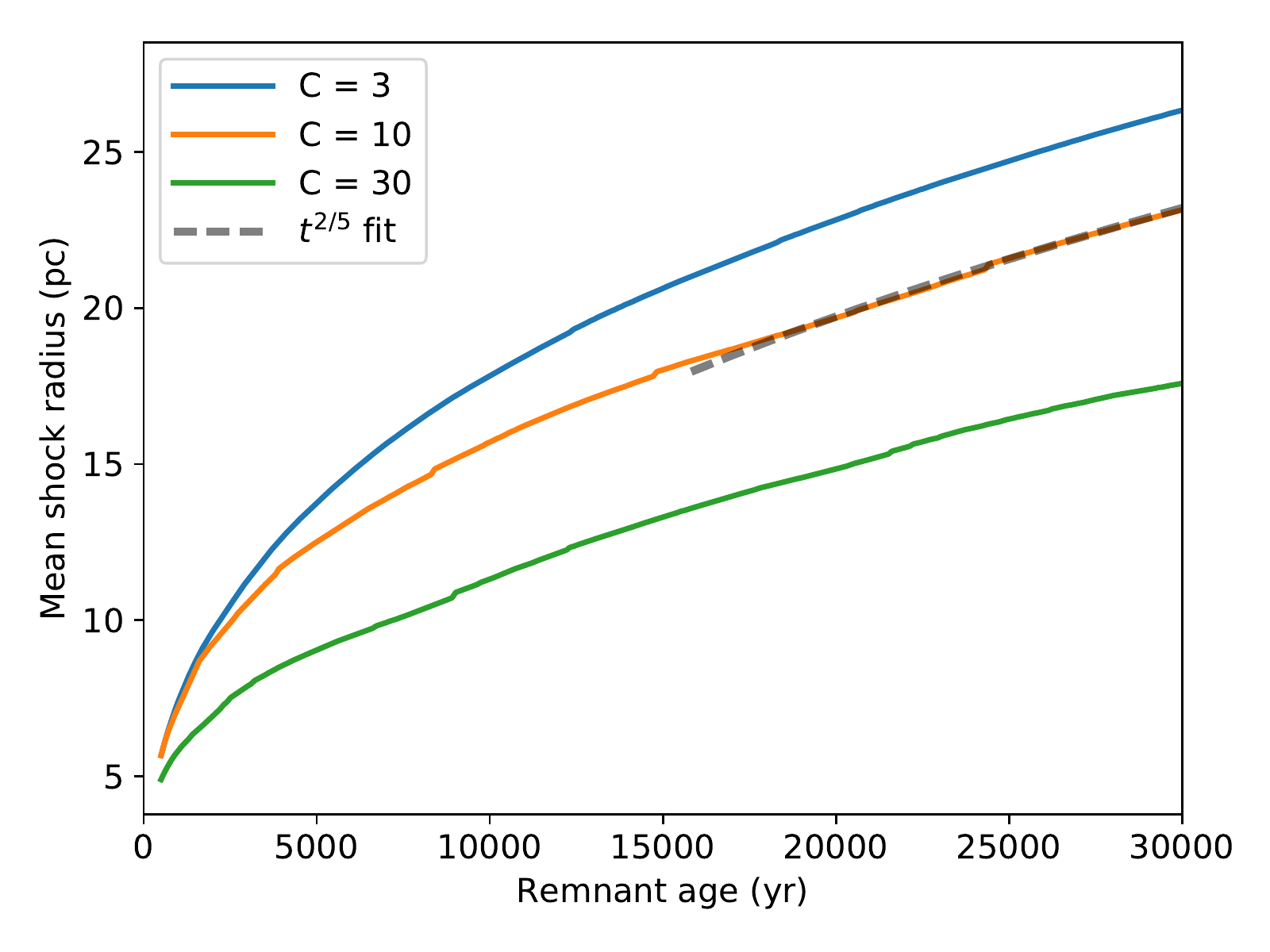}
    \caption{Comparison of the evolution of the mean shock radius for
        runs with different values for $C$, the ratio of ISM mass in clouds
        to that in the intercloud medium.  The density of the clouds and the
        intercloud medium is the same for all of the runs, $25$ cm$^{-3}$
        and $0.25$ cm$^{-3}$, respectively.  Only the filling factor of the
        clouds is varied.  As expected larger $C$ values lead to slower shock
        expansion. The dashed curve labeled ``$t^{2/5}$ fit'' is a fit to the
        shock radius evolution assuming that $R_s \propto t^{2/5}$ for the $C
        = 10$ case, where the fitting was done only for the range in $t$ as
        plotted.  This shows that the evolution approaches the $t^{2/5}$ law
        found by WL. For this case an effective density, i.e.\ the density
        that the corresponding Sedov-Taylor expansion would require, is 0.455
        cm$^{-3}$.\label{fig:rs_C}}
\end{figure}

For our calculations we do find that the shock evolution depends to some
degree on the particular cloud realization used.  This is illustrated in
Figure \ref{fig:rs_evol} which shows the shock evolution for several different
cloud realizations.  For most of our runs we have used only half of the 2D
volume, $z \geq 0$.  The figure shows that the shock evolution for different
runs using the 2D half volume differ though the variations are at about the
10\% level. We have tested the effects of using the 2D half volume by doing a
full volume run, shown as the red dash-dotted line in Figure
\ref{fig:rs_evol}.  We find no significant difference in that run as compared
with other runs with the same $C$ value.  We have also compared with a 3D runs
(using one octant of the space) done in cartesian coordinates.  Again, as
demonstrated in Figure \ref{fig:rs_evol} we find no significant difference
from the 2D half volume runs.  Because of the computational and visualization
demands of the 3D runs, they were done with only 5 levels of grid refinement
rather than the 6 levels used for the 2D runs.  We have found, by doing 2D runs
at 5 and 6 levels of refinement, that while there are minor differences in the
density distributions, with more filamentary structure visible in the higher
resolution runs, the overall evolution of the remnants are nearly identical.
Thermal conduction aids in this respect since small scale details tend to be
smoothed out when it is included.

The expansion law for the runs shown in Figure \ref{fig:rs_evol} is generally
not far from the WL $t^{2/5}$ value, though again there are significant
variations around that.  One way to test that is to compute $v_s t/r_s$, which
for a $t^{2/5}$ expansion law should give 2/5.  Doing this involves taking
numerical derivatives of the shock radius expansion curve which inevitably
results in a noisy curve.  In general we find that the expansion law is close
to but slightly below this value on average, though with excursions above and
below it. 

In Figure \ref{fig:rs_C} we show the shock radius evolution for runs with
different $C$ values.  Since the mean density in the medium increases with $C$
it is expected that the shock expansion rate should decrease as $C$
increases. In the results of WL, the expansion rate is characterized via the
parameter $K$ (their equation 8) where 
\begin{equation}
    r_s = \left[\frac{25(\gamma + 1) K E}{16 \pi \rho_1}\right]^{1/5} t^{2/5},
    \label{eq:rs}
\end{equation}
$r_s$ is the shock radius, $E$ is the explosion energy, $\gamma = 5/3$,
and $\rho_1$ is the ambient intercloud density (0.25 cm$^{-3}$ for our runs).
WL lists values for $K/K(\mathrm{Sedov})$ as a function of $\tau$, the ratio
of cloud evaporation time to remnant age.  $K(\mathrm{Sedov}) = 1.528$ is the
$K$ value for the case of no clouds. We can invert equation \ref{eq:rs} to
determine the effective value of K for our runs.  Doing this we find that K
varies considerably over a typical run though it flattens at late times. For
most of our runs, K defined this way starts high, $> 2$, but ends near 0.9.
The WL model predicts, given a value of $\tau$ of $\sim 5 - 10$ (see
discussion below) and $C = 10$, $K \sim 0.816 - 1.149$.  This is in fairly
good agreement with our results.  K can also be alternatively be defined using
WL's equation (6) as 
\begin{equation}
    K = \frac{4 \pi \rho_1 r_s^3 v_s^2}{(\gamma + 1)E}.
\end{equation}
With this definition, we find more stable values for K, which still fall
roughly in the predicted range.

\subsection{Cloud Evaporation Rate}
Within the WL model the value of $\tau$, the ratio of evaporation timescale to
remnant age, is treated as a free parameter.  In our numerical work, this is
not the case, since given the initial conditions the cloud evaporation
timescale will follow from the physics, including the temperature of the
surrounding hot gas and the complex heat flow patterns that develop as the
clouds are both evaporated and disrupted by shear flows. In addition, clouds
close to the explosion center can be heated enough by being shocked that they
could be considered part of the hot gas, though still overdense.  However, as
we discuss below, such clouds are overpressured compared to the surrounding
medium and so they expand and cool at later times in the remnant evolution.

Clouds that are farther from the center do not get heated to high temperatures
by the expanding shock but are subject to shear flows and thermal conduction.
However, even for these clouds, characterizing the evaporation rate is not
entirely straightforward since gas exists at a range of temperatures and
densities within the remnant at any given time. As a result defining which gas
has been evaporated is not entirely clear cut.

\begin{figure}[ht!]
    \includegraphics[width=0.49\textwidth]{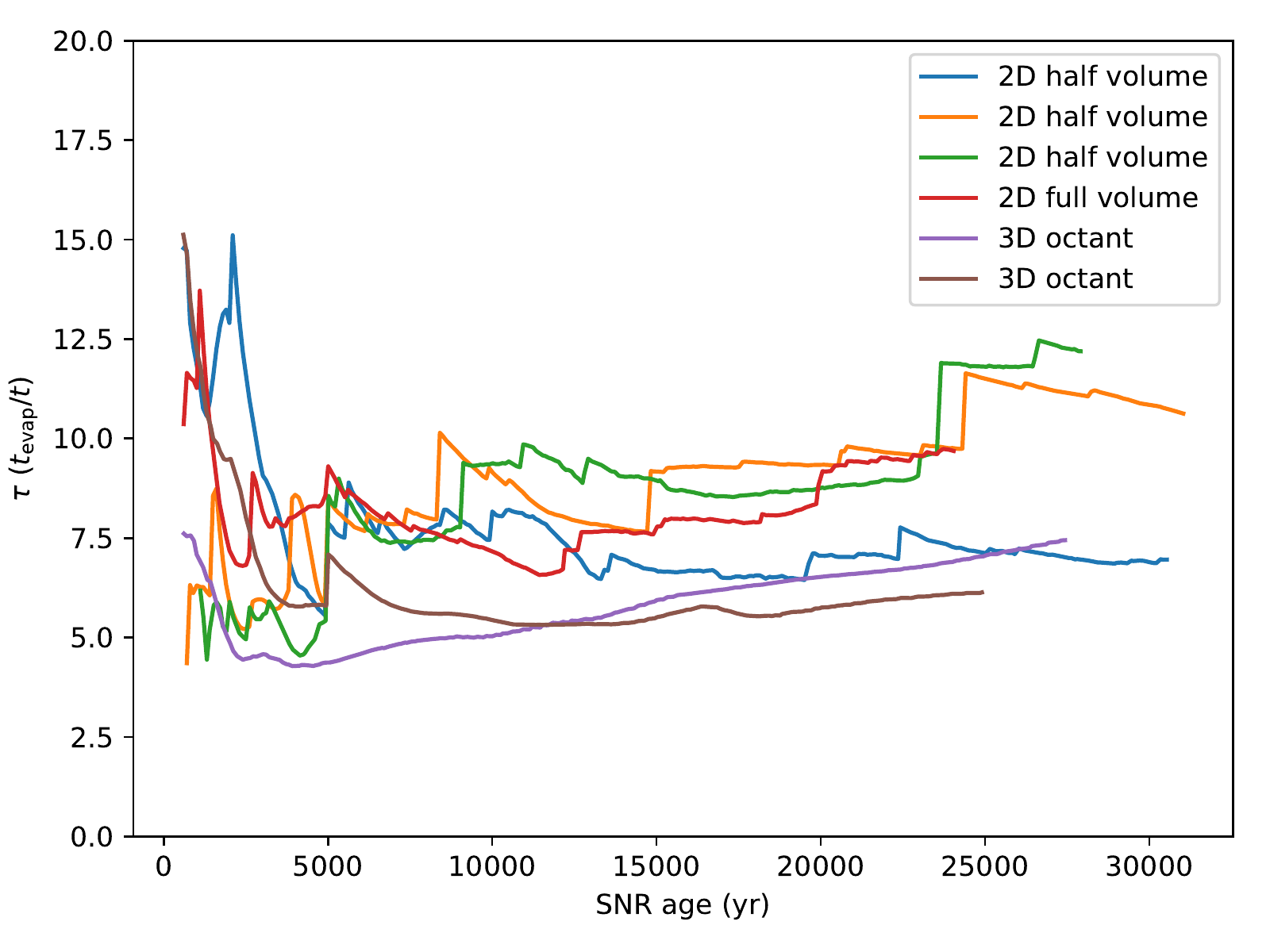}
    \caption{Evaporation timescale factor, $\tau$, of WL vs.\ remnant age.  In
        the WL model $\tau$ is a constant, but here we see that it varies
        considerably over the age of our simulated remnants.  The 3D runs show
        smoother behavior because more clouds are sampled at any given age.
        Here $\tau$ is calculated using the density criterion for deciding
        whether gas is intercloud or cloud gas with $n < 2.5$ cm$^{-3}$ as the
        dividing line.  All the runs plotted here have $C$ set to 10.
    \label{fig:tevap}}
\end{figure}

We have tried a variety of approaches for determining the effective
evaporation timescale for our simulated remnants.  One approach is to use a
density criterion to decide what is cloud material and what is evaporated.
In our case we defined the mass in clouds as the mass of gas with density
above a 2.5 cm$^{-3}$, which is the geometric mean of the density of
intercloud medium, 0.25 \cc, and the cloud density, 25 \cc.  The mass of gas
evaporated from clouds at any given point in the remnant's evolution is then
the mass that was in clouds initially within the shock volume minus the mass
in clouds at the current time. The average evaporation rate is then the
evaporated mass divided by the remnant age.  Connecting this evaporation rate
to that predicted in the WL model is not entirely straightforward however,
since they did not tally that quantity.  Instead we can connect the intercloud
mass, that is all the mass inside the shock that is not in a cloud, with the
total integrated mass in the WL model since the density, $\rho$, or scaled
density, $g = \rho/\rho_s$ stands for all the non-cloud gas in the model.
Thus we compare the intercloud gas in a simulation with
\begin{equation}
    M_x = \int_0^{r_s} 4 \pi r^2 \rho\,dr = 4\pi r_s^3 \rho_s \int_0^1 g(x)
    x^2 \,dx,
\end{equation}
(from WL) where $g \equiv \rho/\rho_s$ is derived by integrating the set of
equations in WL for given values of $\tau$ and $C$.  To find the value of
$\tau$ corresponding to a particular time for a simulated remnant we use the
actual value for $C$, calculated as the initial ratio of mass in clouds to
that in the intercloud medium within the shock volume.  We then do a search,
using a root finding method, wherein we calculate the values of $M_x$ given 
values of $\tau$ until we match the value for the simulated remnant.  This is
the method used to produce Figure \ref{fig:tevap}.

\begin{figure}[ht!]
    \includegraphics[width=0.49\textwidth]{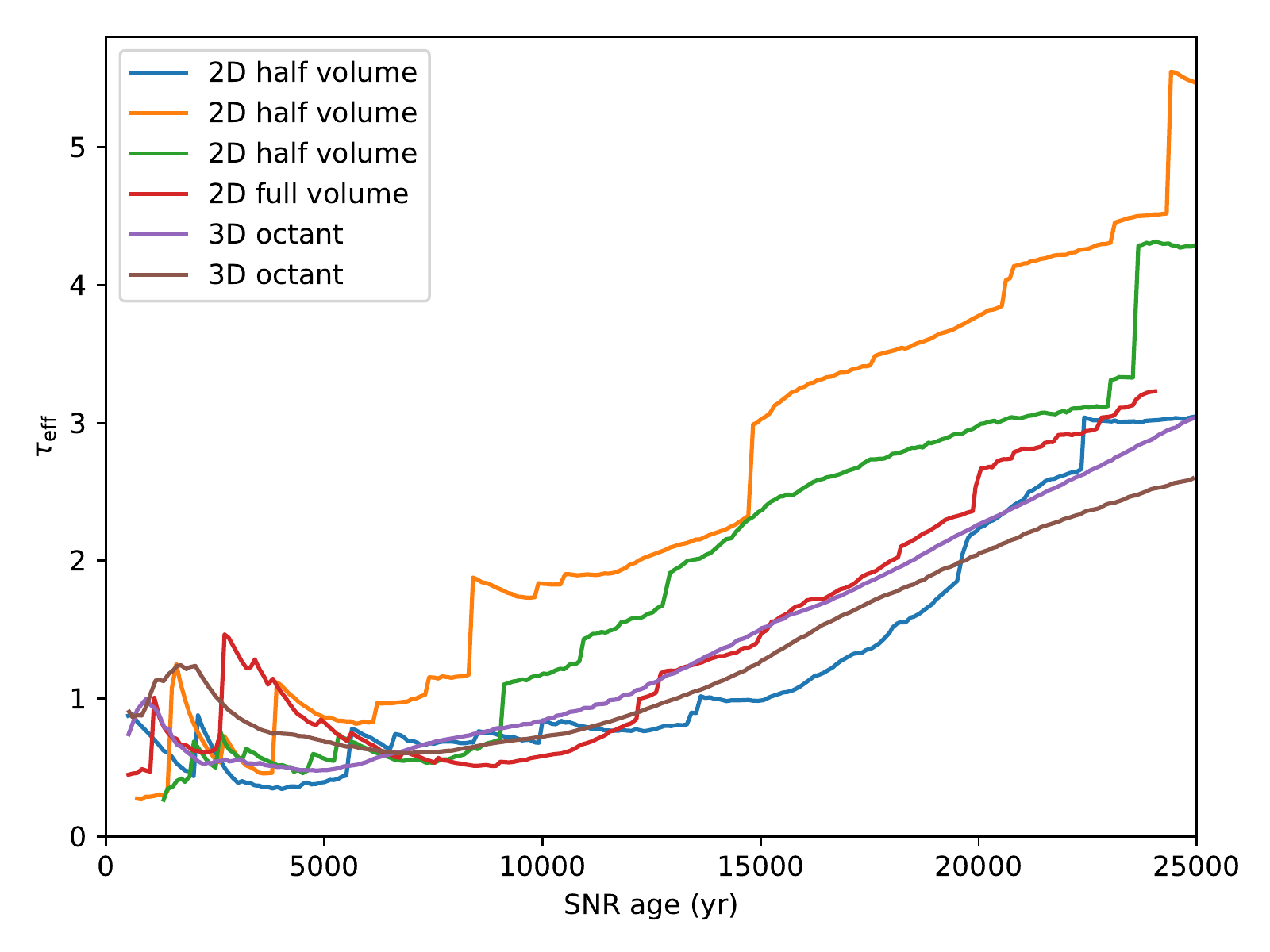}
    \caption{Evaporation timescale factor, $\tau$, of WL vs.\ remnant age.  In
        this case we use the temperature criterion to differentiate between
        cloud and intercloud gas with $T > 10^5$ K as the dividing line.
        This method ensures that the X-ray emitting gas mass (though not the
        luminosity) for the corresponding WL model matches that for the
        simulated remnant. The $\tau$ values calculated this way are
        substantially below those in Figure \ref{fig:tevap}, primarily because
        of hot, but still dense, gas generated by shocking clouds close to the
        explosion center.  Such gas is counted as cloud gas (and thus not
        evaporated) by the method of the previous figure, but should radiate
        X-rays.
    \label{fig:tau_eff}}
\end{figure}

Another approach is to calculate the mass of X-ray emitting gas, essentially
all gas in the remnant that is hotter than 10$^5$ K, and match that with the
X-ray emitting mass, $M_x$, for a given WL model.  That approach leads to
values of $\tau$ that are considerably lower than for the density based
criterion, mostly because of the production of hot, dense gas via shock
heating.  The results for $\tau$ using the temperature based criterion
are shown in Figure \ref{fig:tau_eff}.

The status of the shocked cloud gas is somewhat ambiguous in the model,
however since it is overpressured when shocked and some of it re-expands and
cools adiabatically at later times below X-ray emitting temperatures. For
clouds shocked by slower shocks, only a surface layer is raised to X-ray
emitting temperatures.  Thus, at later times, the removal of material from
clouds is primarily by thermal conduction and cloud evaporation as discussed
by WL does apply.

We expect that the evaporation timescale for the remnant as a whole should
depend in a similar way to that for individual clouds \citep[though
see][regarding the effect on evaporation rate of a collection of
clouds]{Balbus_1985}.  For ``classical'' evaporation, i.e.\ cloud evaporation
when the heat flux is far from saturation, the evaporation timescale goes as
$t_\mathrm{evap} \propto R_\mathrm{cl}^2 n_\mathrm{cl} T_h^{-5/2}$ where
$R_\mathrm{cl}$ is the cloud radius, $n_\mathrm{cl}$ is the cloud density and
$T_h$ is the temperature of the hot gas far from the cloud.  In the case of
highly saturated heat flux, which is the more typical condition in SNRs during
the non-radiative phase, we get $t_\mathrm{evap} \propto R_\mathrm{cl}^{7/6}
n_\mathrm{cl} P^{-5/6}$, where $P$ is the thermal pressure in the hot gas.
Since, ignoring the effects of shocks, the density in the clouds is expected
to be roughly constant and, as mentioned above, the pressure should decrease
as $t^{-6/5}$, the evaporation timescale should increase proportional to $t$.
From this we would expect the WL parameter $\tau \equiv t_\mathrm{evap}/t$
to be roughly constant and to only depend on the density and size of the
clouds.

For our assumptions regarding mean mass per particle we find
\begin{equation}
    t_\mathrm{class} = 3.02\times10^{14}\frac{R_\mathrm{cl}^2
    n_\mathrm{cl}}{T_h^{5/2}}\;\mathrm{Myr}
\end{equation}
and
\begin{equation}
    t_\mathrm{sat} = 1.48\times10^3\frac{R_\mathrm{cl}^{7/6}
    n_\mathrm{cl}}{P_h^{5/6}}\;\mathrm{Myr}
\end{equation}
where $R_\mathrm{cl}$ is the cloud radius in pc, $n_\mathrm{cl}$ is the cloud
density in cm$^{-3}$, $T_h$ is the temperature of the surrounding hot gas and
$P_h$ is the thermal pressure (presumed to be roughly equal in the cloud and
hot gas) in cm$^{-3}$ K.  We find that with these values, it is not typical to
get values for $\tau$ less than one.  These expressions also make it clear
that the primary determinant of the $\tau$ value is the cloud size
distribution.  We have confirmed this by doing runs with smaller clouds, using
0.2 and 5.0 pc as the lower and upper size limits, but with the same volume
filling factor (and thus $C$ value).  As expected we find that the clouds
evaporate faster, leading to smaller values of $\tau$.

While our discussion above indicates that $\tau$ should be roughly constant,
we find that it evolves considerably over the course of SNR evolution.  This
is not too surprising since, especially for the 2D evolution, a small number
of clouds is encountered and the location and size of the clouds affects the
evaporation rate.  This is demonstrated in Figures \ref{fig:tevap} and
\ref{fig:tau_eff} where we plot $\tau$, calculated two different ways, as a
function of remnant age for a variety of our runs.  The differences
between the 2D runs can be attributed to the differences in the particular
cloud realization for the runs with some with larger clouds placed closer to
the origin and the spacing of clouds differing.  As can be seen, the 3D runs
show less variation between themselves and over time.  This can be understood
as deriving from the larger number of clouds encountered and thus more
complete sampling of the cloud size distribution.  A full volume 3D run would
likely show an even smoother variation though the finite cloud volume would
still be expected to create variations over the course of the cloud evolution.
There is some indication here that there is faster evaporation (smaller
$\tau$) in the 3D simulations than in the 2D simulations, which could be
understood as being caused by the larger surface area to volume ratio
effectively for the spherical clouds in 3D than for the toroids under
cylindrical symmetry.  Nevertheless, the shock expansion does not show any
marked difference between 2D and 3D cases nor does the calculated X-ray
emission as we discuss in the next section.

\begin{figure*}[ht!]
    \includegraphics[width=0.98\textwidth]{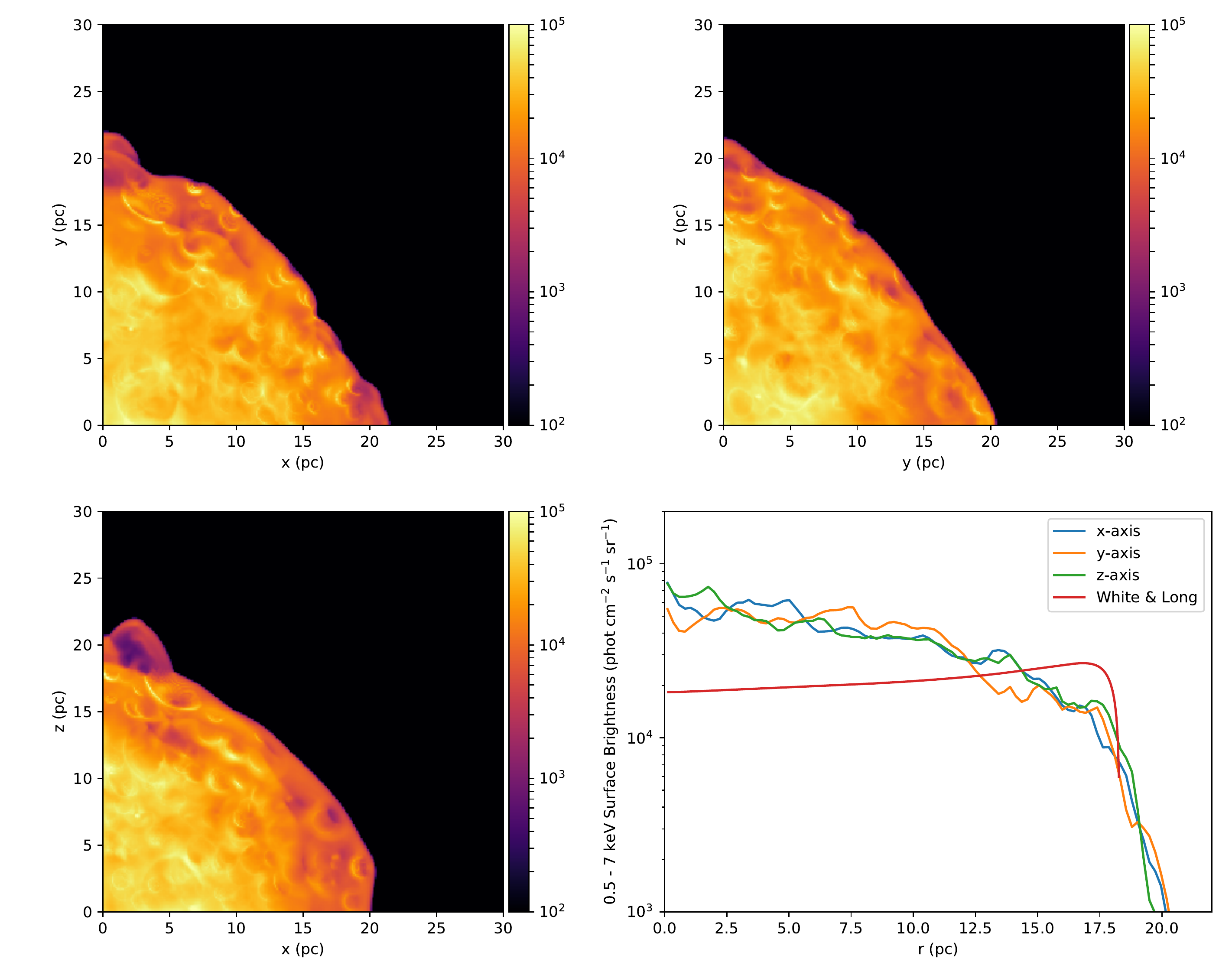}
    \caption{Brightness profiles for a 3D run in the $0.5 - 7$ keV band.  The
        images are created by integrating along a given axis while the lower
        right line plot is radially averaged in addition. The emissivity
        assumes collisional ionization equilibrium and uses APEC
        \citep{Foster_etal_2012,Smith_etal_2001} to calculate the photon
        emissivity as a function of temperature. The age of the remnant is
        $1.5\times10^4$ yr. For the White \& Long curve we used the actual
        values determined at the given time for the values for $C$ and $\tau$
        which are 7 and 9 respectively.
    \label{fig:bprof}}
\end{figure*}

\begin{figure*}[ht!]
    \includegraphics[width=0.98\textwidth]{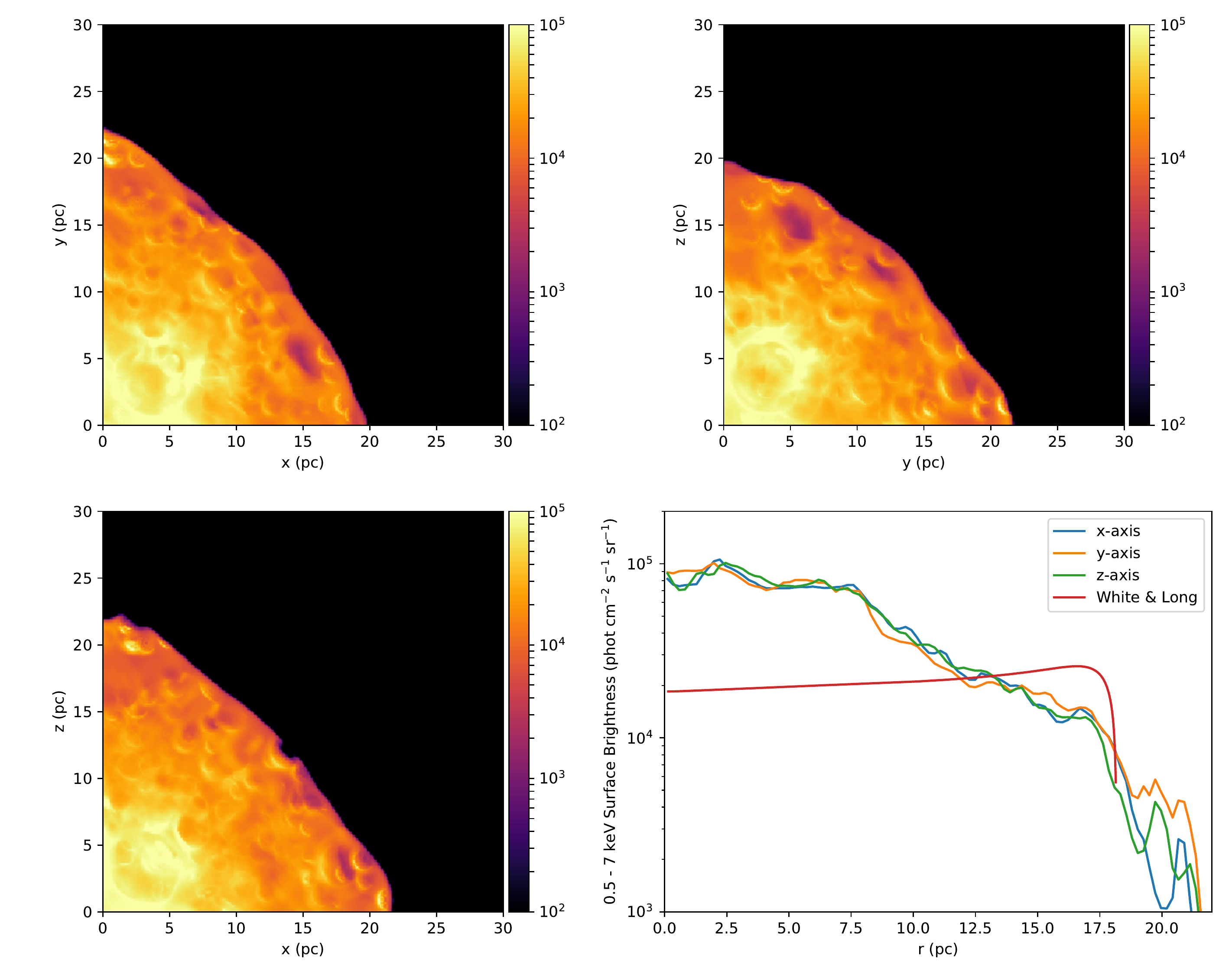}
    \caption{Same as Figure \ref{fig:bprof} except for a run with a different
        cloud realization.  The differences in the brightness profiles
        illustrate the level of variation caused by differences in the random
        distribution of clouds in the ambient medium.
    \label{fig:bprof2}}
\end{figure*}

\subsection{X-ray Emission Distribution}
\begin{figure}[ht!]
    \includegraphics[width=0.49\textwidth]{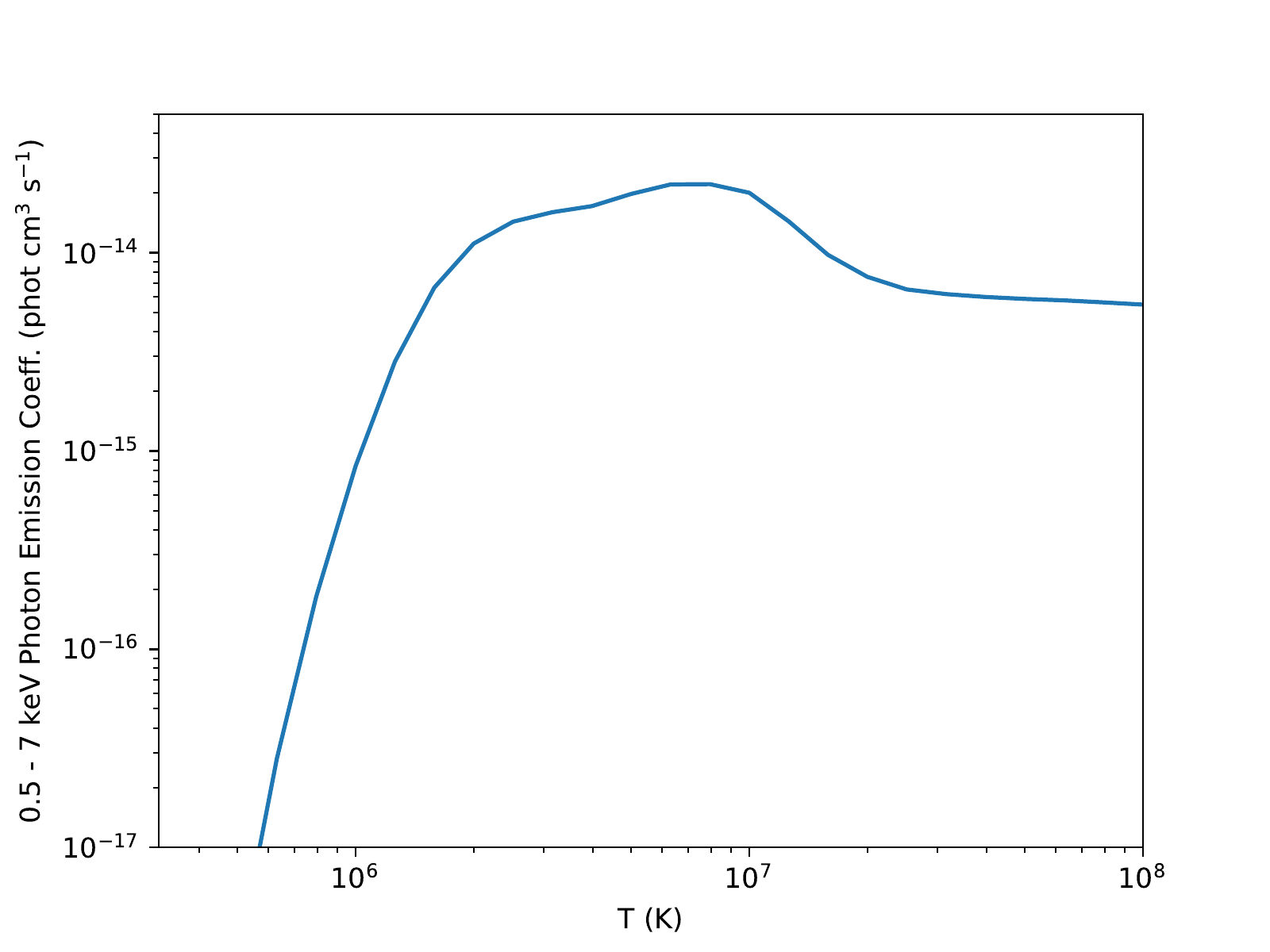}
    \caption{Emission coefficient for optically thin radiative cooling by gas
        in collisional ionization equilibrium.  This coefficient is for
        photons emitted in the $0.5 - 7$ keV band and includes both line and
        continuum emission.
    \label{fig:APEC_emis}}
\end{figure}

One of the principal motivations for the WL model was to explain the
appearance of SNRs with centrally peaked thermal X-ray emission. WL showed
that they could achieve a range of different X-ray emission distributions
depending on the choice of their parameters $C$ and $\tau$. From their Figure
4 it can be seen that centrally peaked X-ray emission requires values of $C
\gtrsim 20$, depending on the value of $\tau$.  For $\tau = 2$ (WL Figure 4a)
none of the emission profiles is truly centrally peaked.  With $\tau = 10$ (WL
Fig. 4b) centrally peaked emission is achieved, though only for fairly high
$C$.  For our evaluation of the emission in our simulated remnants we have
chosen to use a more realistic emissivity than the constant $\Lambda_x =
10^{-22}$ ergs cm$^{3}$ s$^{-1}$ used by WL (note that in the units for this
coefficient, the exponent for cm in WL was mistakenly shown as -3 rather than
3).  Here we use a photon emissivity for optically thin emission in the $0.5 -
7$ keV band (see Figure \ref{fig:APEC_emis}) as calculated using APEC
\citep{Foster_etal_2012,Smith_etal_2001} with the assumption of collisional
ionization equilibrium (CIE). We discuss the assumption of CIE below.

This choice of emission coefficient covers a typical X-ray observation energy
range, e.g. with Chandra. From Figure \ref{fig:APEC_emis} it is clear that
this emissivity weights the emission profile strongly toward gas in the
temperature range of $\sim 10^{6.5} - 10^7$.  A detector with sensitivity at
lower energies could see the emission from lower temperature gas generated at
the boundaries of the clouds which is expected to be strong. The lower right
panels of Figures \ref{fig:bprof} and \ref{fig:bprof2} show a
comparison of the WL profile that corresponds with the $\tau$ and $C$ values
at that time for each run.  Note that here we use the effective value for $C$
rather than the mean value.  By this we mean that we use $C =
m_\mathrm{cl}/m_\mathrm{ic}$ where $m_\mathrm{cl}$ is the cloud mass,
$m_\mathrm{ic}$ is the intercloud mass and both are for the initial medium
contained within the current volume of the shock.  This will tend to make the
effective $C$ smaller than the mean value over the whole medium because the
shock is slowed where it encounters clouds, though in practice we have found
this effect to be fairly minor.  The same APEC emissivity was used to
generate the WL profile as for those from the simulations.  The relative
flatness of the WL profile is caused by the lower density and higher
temperature near the center as compared with the simulations (see Figure
\ref{fig:prof}).

\begin{figure}[ht!]
    \includegraphics[width=0.49\textwidth]{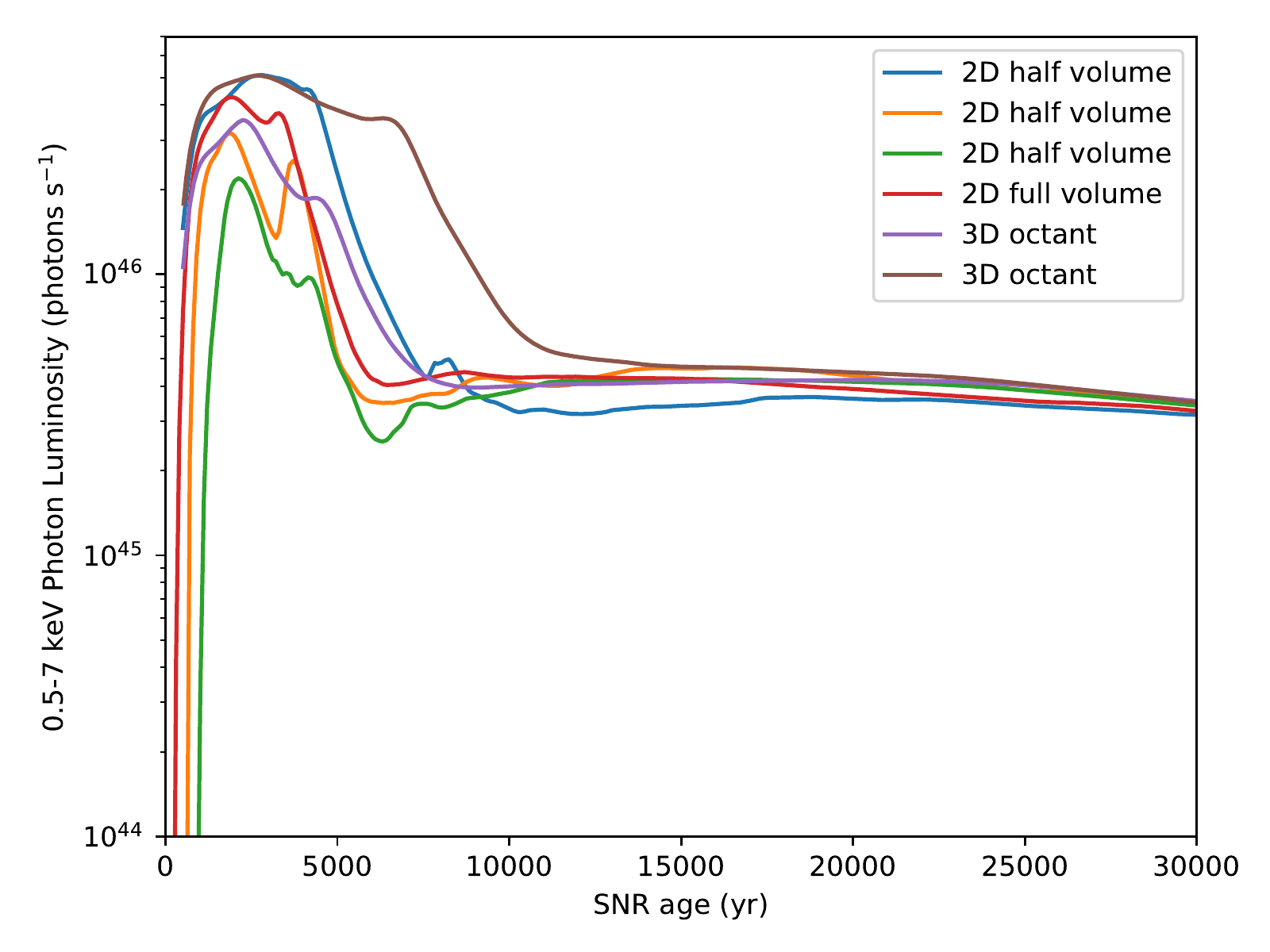}
    \caption{Evolution of the $0.5 -7$ keV photon luminosity in our
        simulations.  At early times the luminosity is low because the very
        high gas temperature and because few clouds have been encountered.
        The peak in emission comes when the shock encounters clouds and is
        still fast enough that the shock transmitted into the clouds can heat
        gas to X-ray emitting temperatures.  Later cloud evaporation raises
        the hot gas density while expansion lowers the pressure, leading to
        roughly constant luminosity.
    \label{fig:lumin_evol}}
\end{figure}

\begin{figure}[ht!]
    \includegraphics[width=0.49\textwidth]{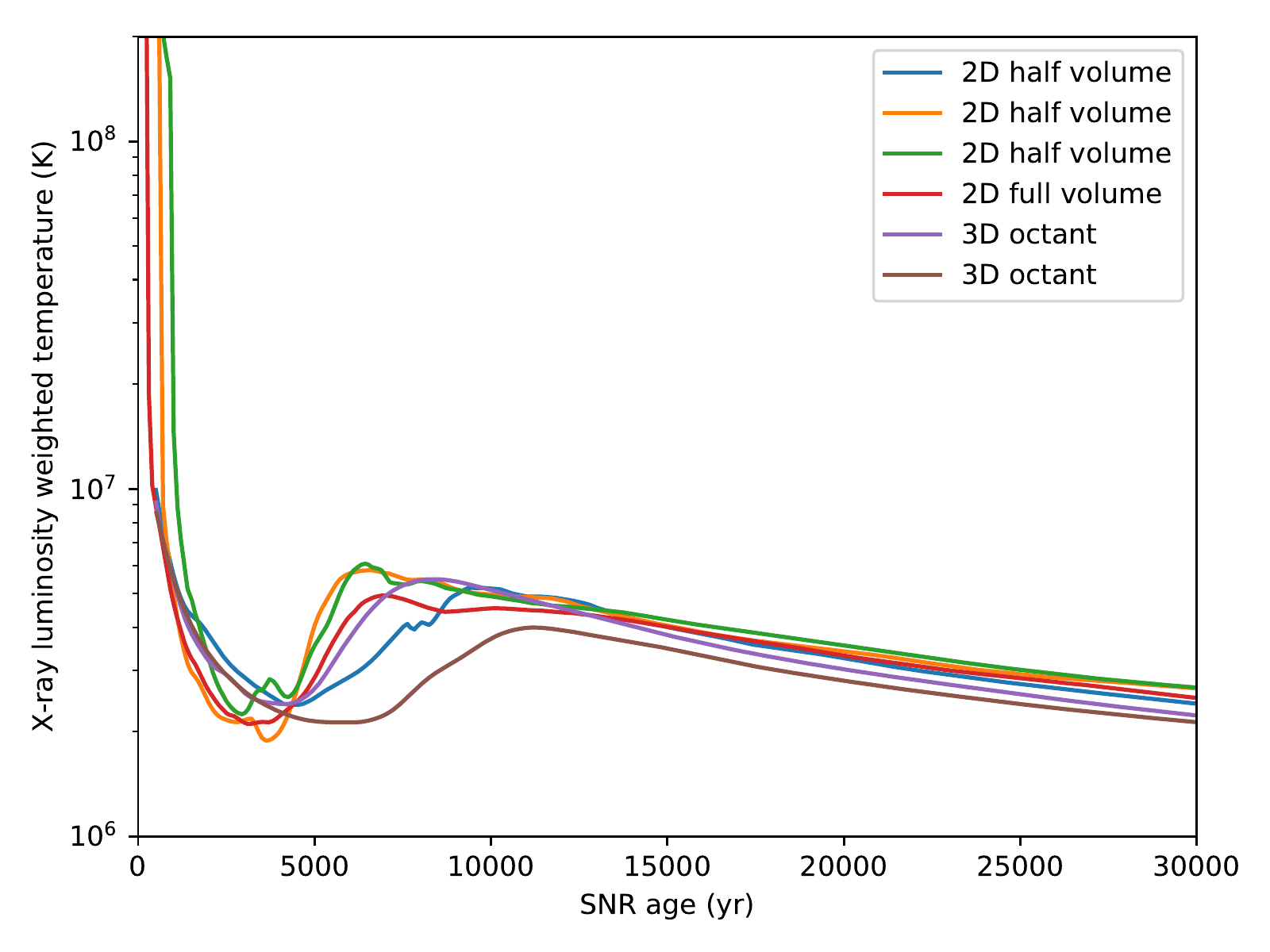}
    \caption{Evolution of the X-ray weighted temperature in simulated
        remnants.  At the time of peak luminosity, much of the emission is
        coming from relatively low temperature shocked cloud gas, which leads
        to low temperature emission. At later times more of the emission comes
        from the hot intercloud gas which is hotter.
    \label{fig:Tlum_evol}}
\end{figure}

The evolution of the $0.5 - 7$ keV photon luminosity is shown in Figure
\ref{fig:lumin_evol} for several simulations with $C = 10$. The
emissivity-weighted temperature is shown in Figure \ref{fig:Tlum_evol}. The
luminosity starts low both because the remnant volume is low (i.e. the shock
radius is small) and because the temperature is very high, above the
temperature range of highest emission efficiency in the band (see Figure
\ref{fig:APEC_emis}).  (Note that because we do not include ejecta mass or
extra circumstellar material, the very early emission is not expected to be
accurately modeled.) The luminosity then reaches a peak at an age of $\sim
2000$ yr.  This peak is associated with a time when the SNR encounters its
first cloud. The shock is fast enough at this point (in our runs) to heat the
entire cloud to X-ray emitting temperatures.  Later the clouds close to the
explosion site re-expand and cool (adiabatically since we have not included
radiative cooling). This is because the shocked clouds are overpressured
relative to the intercloud medium and also the pressure in the remnant is
decreasing as it expands.  An example of this can be seen in Figure
\ref{fig:slice3D} where a hollow looking cloud can be seen which has
re-expanded and cooled.  These clouds cool enough that they fall below X-ray
emitting temperatures, i.e. $T \lesssim 10^5$ K.  For clouds farther from the
origin, only their inward facing edges get heated sufficiently to produce
X-rays (see the arcs in Figure \ref{fig:bprof}). Cloud material at those later
times is continually being evaporated off the clouds, which increases the
density of the hot gas, but at the same time the expansion of the remnant
cools the gas.  The net result, as can be seen from the figure, is an almost
flat or slightly decreasing X-ray photon luminosity.  This contrasts with the
behavior predicted by WL (their eq. 21) which is for the luminosity to
increase in proportion to the remnant volume.  The evolution of the emissivity
weighted temperature of the emission has a nearly inverted profile compared
with the luminosity evolution.  That can be explained by the fact that the
shocked cloud gas is relatively cool, though dense, so when that gas is hot
enough to emit X-rays, the emissivity-weighted temperature is low.  As a
remnant transitions to the phase where most of the emission is coming from the
lower density and hotter gas, the emission temperature increases and then
flattens out.

These differences in luminosity evolution are some of the most substantial
differences between our calculated results and the WL model.  The domination of
the emission by the shocked cloud gas shows that the MM remnant's brightness
distribution, especially at early times, does not depend as much on
evaporated mass as on the shock interactions of the blast wave with the
clouds.  In this sense, the value of $\tau$ derived based on density, as shown
in Figure \ref{fig:tevap} is not such an important parameter, though it is
consistent with our expectations for the evaporation timescale, including the
fact that it is roughly constant during remnant evolution.  Instead the
effective $\tau$, shown in Figure \ref{fig:tau_eff}, where the temperature of
the gas is the criterion for differentiating cloud gas from intercloud gas
is a better measure for understanding the X-ray luminosity evolution, though
given that some of that gas later cools via expansion, this $\tau$ is not
a good measure of gas that has been thermally evaporated.

\section{Discussion}

There are a number of additional processes that could affect SNR-cloud
interaction in the ISM.  Chief among these is radiative cooling.  WL did not
include cooling and we do not include it in our results presented here, though
we intend to explore it in future work.  In preliminary calculations we have
found that the shape and size of shocked clouds is strongly affected, but the
overall evolution of the remnant, for ages early enough that the remnant has
not yet gone radiative as a whole, is not substantially changed.
\citet{Korolev_etal_2015} have recently studied the longer term evolution of
SNR evolution in a cloudy medium including radiative cooling but not thermal
conduction with 2D numerical simulations.  

If radiative cooling is important, then non-equilibrium ionization should be
taken into account.  In SNRs in general and for shocks into clouds in
particular, we expect the post shock gas to be underionized and far from
CIE.  This is also true for gas that is evaporated from the entrained clouds
\citep{Ballet_etal_1986,Slavin_1989}. It is also expected that the hot gas
that was created early in the remnant's lifetime will be overionized after the
remnant has expanded and cooled.  This effect will be enhanced by the presence
of thermal conduction which draws thermal energy away from the hot interior
toward the cooler outer parts of the remnant. In addition there could be
turbulent mixing in the medium that would combine hot and cold gas producing
yet another type of non-equilibrium ionization state \citep{Slavin_etal_1993}.
All of these effects can have potentially large impacts on the emitted X-ray
spectrum if the ionization is sufficiently far from CIE.  These effects will
need to be taken into account in order to make quantitative predictions for
the emission spectrum for SNRs evolving in a cloudy ISM. We are currently in
the process of improving the treatment of non-equilibrium ionization by FLASH,
which will then allow efficient and accurate calculations of NEI effects.

Another possibly important influence on remnant evolution is the magnetic
field.  The field can affect the overall expansion of the remnant as well as
the evaporation of clouds in the remnant.  The anisotropic thermal conduction
within the remnant when the clouds are threaded by the field is likely to have
complex effects which deserve to be studied in detail.  Our aim in the current
study has been to examine SNR evolution under conditions similar to those
intended in WL and so we do not include the magnetic field at this stage.
Finally, we do not include heating via photoionization.  This could be
important in the post-shock regions of radiative shocks but is likely a
minor effect for the young to middle-aged remnants evolving in the low
density, warm ISM such as we examine in this work.

\section{Conclusions}
The WL model was put forward in an attempt to explain the mixed morphology
class of SNRs as being caused by the evaporation of dense clouds that are
overrun by the blast wave.  We have presented numerical hydrodynamical models
of SNRs evolving in a cloudy medium including thermal conduction in order to
test whether the results of WL hold up when more of the physics is included.
We find that some of their results are broadly confirmed including: 
remnants expand with roughly the same $t^{2/5}$ power law dependence on age as
a standard Sedov-Taylor expansion, 2) the cloud evaporation timescale
increases roughly in proportion to the remnant age (i.e.\ $\tau$ is roughly
constant), and 3) the presence of clouds causes the remnant X-ray brightness
to be centrally peaked.  However, the lack of the inclusion of the dynamical
effects of cloud-shock interaction and thermal conduction within the hot gas
that are present in multidimensional calculations lead to substantial
differences from the results of WL.  Probably the largest difference is with
the predicted X-ray luminosity evolution which, rather than rising in
proportion to the shocked volume, has an early peak and then flattens to a
nearly constant level.  This can be traced to the shock heating of the clouds
close to the explosion site, which provides the peak, and later to the
expansion cooling of the shocked clouds along with a transition to evaporative
mass loss as the dominant mass loss mechanism for the clouds.  The differences
in temperature and density distribution in the remnants caused by thermal
conduction within the hot gas and the shocking of the clouds leads to a
flat X-ray brightness distribution in the remnants.  For certain combinations
of parameters, the brightness profile for WL models can match that from our
models, though the parameters do not correspond to the physics, i.e.
evaporation timescale and cloud-to-intercloud mass ratio, actually present in
the simulation.  In general we find that the X-ray emission weighted
temperature is lower for our models than for WL models because of the
inclusion of lower temperature hot gas associated with the clouds, either
shocked or evaporated gas.

Our results point to important differences from the WL model for SNRs evolving
in a cloudy ISM which could have implications for the interpretation of SNR
observations, particularly in the X-rays.  However, to make more robust
predictions, particularly for the X-ray spectrum, will require the inclusion
of more physical processes in future simulations.  In particular radiative
cooling and non-equilibrium ionization could substantially affect the observed
spectrum of SNRs such as those we have modeled.  We have begun to look at the
effects of these processes and those calculations will allow us to make
detailed predictions for the spectra of mixed morphology SNRs in future work.

\acknowledgements
Software used in this work (FLASH) was in part developed by the DOE NNSA-ASC
OASCR Flash Center at the University of Chicago.  This work was supported by
NASA Astrophysics Theory Program grant NNX12AC70G.

\software{FLASH \citep{Fryxell_etal_2010}}

\appendix

\section{FLASH parameters for the use of thermal conduction}

The \texttt{Diffuse} module of FLASH uses a number of parameters that govern
how thermal conduction is calculated.  Both standard Spitzer conductivity,
$\kappa \propto T^{5/2}$, and saturated conductivity \citep{McKee+Cowie_1977}
$\kappa \propto \rho c^3$ are supported with a smooth transition from
``classical'' to saturated governed by the saturation parameter $\sigma$.  The
parameter values that we have used in the \texttt{flash.par} files for runs
with thermal conduction included are listed in table \ref{tab:condpar}. The
particular unit that we used is included via the switch
\texttt{-unit=physics/Diffuse/DiffuseMain/Unsplit} to the \texttt{setup}
command. In addition, to use the power law conductivity, we added
\texttt{REQUIRES physics/materialProperties/Conductivity/
ConductivityMain/PowerLaw} to the \texttt{Config} file in our simulations
directory.  With these settings, electron thermal conduction is used with a
power law conductivity, $\kappa = a T^{5/2}$ modified by the saturation limit
from \citet{Cowie+McKee_1977} and using the harmonic mean weighting (chosen
through the \texttt{diff\_eleFlMode} parameter) as in
\citet{Balbus+McKee_1982}, $q = (1/q_\mathrm{class} + 1/q_\mathrm{sat})^{-1}$,
where $q_\mathrm{class}$ is the ``classical'' Spitzer heat conduction and
$q_\mathrm{sat}$ is the ``saturated'' conduction, which corresponds to the
heat flux being transported at the maximum rate possible by the electrons,
$q_\mathrm{sat} = 5\phi_s \rho c^3$. \citet{Balbus+McKee_1982} argue that
$\phi_s \approx 0.3$ (though with substantial uncertainty) and we use that
value in this work. Here $a$ is the Spitzer conductivity coefficient
\citep{Spitzer_1962}, $1.84\times 10^{-5}/ln\Lambda$, where $\Lambda$ is the
Coulomb logarithm. For conditions that we are exploring, $a \approx 6\times
10^{-7}$. In FLASH the implementation of saturation uses $q_\mathrm{sat}
= \alpha_\mathrm{ele} \rho_\mathrm{ele} c_\mathrm{ele}^3$
effectively, so to use $\phi_s = 0.3$ we set $\alpha_\mathrm{ele} = 0.04491$.
This is derived under the assumption of a fully ionized plasma with a He
abundance of 10\% by number.  These assumptions also lead to setting
\texttt{eos\_singleSpeciesA} to 0.6123. In reality the cooler parts of the
plasma will most likely be partially ionized or nearly neutral, though in
those regions conductivity is very low in any case. The parameter
\texttt{diff\_thetaImplct} sets the scheme for the diffusion solver, where 0.5
is for the Crank-Nicholson method, 0.0 for fully explicit and 1.0 for fully
implicit.  As indicated in the table, we use the Crank-Nicholson scheme. Since
that scheme is unconditionally stable we set the \texttt{dt\_diff\_factor} to
$10^{10}$ so as to effectively prevent the very restrictive diffusion timestep
constraint from limiting the timestep.

We have tested the thermal conductivity by calculating the steady evaporation
of a spherical cloud in cylindrical symmetry (in 2D) under moderately
saturated conditions.  The resulting mass loss rate and temperature profile
closely matched that predicted by the results of \citet{Dalton+Balbus_1993}
who found analytical solutions for steadily evaporating clouds as a function
of the degree of saturation.  These results give us confidence that thermal
conduction is functioning correctly in the code.

\begin{deluxetable}{ll}
    \tablecaption{Diffusion related parameters settings for FLASH runs with
    conductivity\label{tab:condpar}}
    \tablehead{
        \colhead{parameter name} & \colhead{parameter value} \\
    }
    \startdata
    useConductivity & \texttt{.true.} \\
    useDiffuse & \texttt{.true.} \\
    useDiffuseTherm & \texttt{.true.} \\
    dt\_diff\_factor & 1.E10 \\
    cond\_densityExponent & 0.0 \\
    cond\_temperatureExponent & 2.5 \\
    cond\_K0 & 6.E-7 \\
    diff\_useEleCond & \texttt{.true.} \\
    diff\_eleFlMode & fl\_harmonic \\
    diff\_eleFlCoef & 0.04491 \\
    diff\_thetaImplct & 0.5 \\
    diff\_eleXlBoundaryType & zero-gradient \\
    diff\_eleXrBoundaryType & zero-gradient \\
    diff\_eleYlBoundaryType & zero-gradient \\
    diff\_eleYrBoundaryType & zero-gradient \\
    diff\_eleZlBoundaryType & zero-gradient \\
    diff\_eleZrBoundaryType & zero-gradient \\
    \enddata
\end{deluxetable}

\bibliographystyle{aasjournal}
\bibliography{snr_evol}

\end{document}